# Videos of physics experiments
# A supplementary educational tool for students and teachers

M. Pilakouta[1], K.Mitritsakis[2], E.Fragkedakis[2], C.P. Varsamis[1]

[1] Dpt. of Physics Chemistry and Material Technology, Technological Educational Institute (T.E.I) of Piraeus, Greece, Tel: +302105381553,
mpilak@teipir.gr, cvars@teipir.gr

[2] Dpt. of Automation, Technological Educational Institute (T.E.I) of Piraeus, Greece,
kostas.mit88@gmail.com, Mfragk@gmail.com

The educational use of video and multimedia is increasing rapidly in secondary and higher education across all disciplines. Videos for physics education can be found in many universities and other educational institutions websites all over the world. In the area of experimental physics, the available videos demonstrate mainly physical phenomena or physics experiments and only few of them allow for the quantitative estimation of physical parameters.

In this work, we present characteristic videos of an ongoing project aiming at the development of a collection of educational videos that guide the users – students to measure data and to analyze them in order to calculate physical quantities. These videos can be used for physics teaching, as a demonstration, as a supplementary educational tool for the students' pre lab preparation and also in the physics lab, if the necessary equipment is not available or in case of time consuming measurements.

The pilot use of a video related to the measurement of the lead attenuation coefficient in the lab is presented accompanied by the student assessment. The students favor the use of video mainly as a supplementary tool for their pre lab preparation and not as a substitute of the real experiment.

**1. Introduction**

The growing development of the Information and Communication Technology (ICT) provided valuable tools in educational technology. In particular, videos and multimedia constitute a foundational technology for distance and distributed learning as well as for traditional classroom-based courses. An impressive number of video lectures, instructional videos, demonstrations and other multimedia applications are available nowadays for free in the websites of universities and other educational institutions [1-4].

Learning sites (Academic Earth, videolectures.net) with videos of university lectures and other educational content from Yale, MIT (more than 2,000 courses online), Harvard, Stanford, UC Berkeley, Princeton and others, emerged recently [2-4]. At the lecture capture area, 2011 is considered the most active year until now though the demand for educationally-targeted video archives is still high [5, 6].

In the area of experimental physics, there are many freely available online videos for teaching and learning several concepts of contemporary physics. The majority of these videos demonstrate physical phenomena or physics experiments and only a minority of them, dedicated for motion analysis [7], allow for the quantitative estimation of physical parameters.

In this work, we present firstly three videos of physics experiments that guide the users – students to collect data and elaborate them to calculate physical quantities. These videos can be used for physics teaching, as a demonstration, but also as a supplementary educational tool for students' preparation before the experiment. Moreover, videos may substitute fully or partly the real experiment in the physics lab, in case that the necessary equipment is not available or the measurements are time consuming. The videos, which are a part of an ongoing program aiming at the development of a collection of freely available online videos for teaching and learning experimental physics, were produced in the physics lab of TEI Piraeus with a common digital camcorder, a free video processor and the collaboration of our students. The target group of the videos is first year undergraduates with a medium background in physics and secondary school students.

Finally, a pilot implementation of a video related to the measurement of the lead attenuation coefficient in the physics lab is shortly presented and the student's assessment for this laboratory lesson as well as their attitude towards the use of the video of physics experiments as a supplementary educational tool is discussed.

**2. Videos of physics experiments**

Short videos describing the experimental procedure of the following physics experiments were recorded and processed properly:
- Measurement of fluids viscosity (FV)
- Measurement of sound velocity (SV)
- Gamma ray attenuation and measurement of lead attenuation coefficient, (GRAC)

Each video includes:
- a short introductory part with the relevant theory,
- the experimental procedure based on a specific scenario for each experiment
- the appropriate set of instructions to get the necessary measurements

In the following paragraphs, the descriptions as well as the possible usage of the three videos are discussed.

<u>2.1 Measurement of fluid viscosity</u>

This experiment aims at the determination of the viscosity coefficient η of a fluid. This is achieved by measuring the terminal velocity $u_T$, of a small sphere, with a radius r in the order of mm, falling in a vertical tube containing the fluid by applying the equation [ref8, p.403]:

$$u_T = \frac{2gr^2(\rho_s - \rho_f)}{9n} \tag{1}$$

Where $\rho_s$ and $\rho_f$ denote the density of the sphere and of the fluid respectively and g the gravity acceleration.

The corresponding video (FV) with a duration of 1.20 min consists of three parts. In the first one, the simultaneous motion of two identical small spheres falling in tubes containing machine oil and glycerin (laminar flow) are shown in "slow motion" (Fig. 1). The video starts with the two spheres at the same position and then shows the evolution of the motion in a small distance.

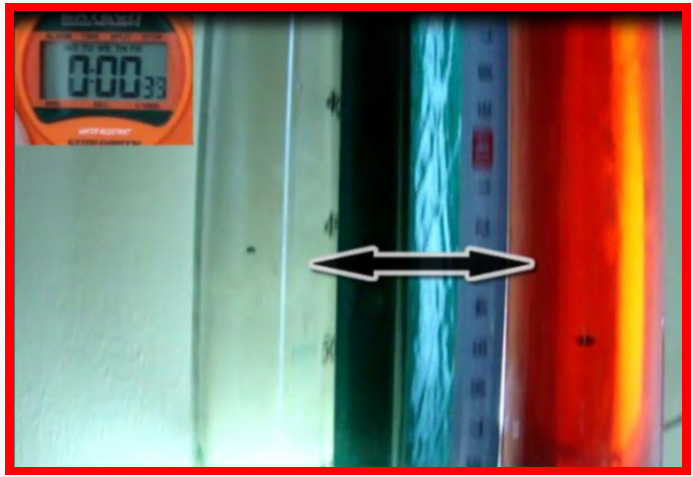

**Figure 1:** *A view of the motion of two identical small spheres, in laminar flow conditions a) in glycerin - left tube and b) machine oil - right tube. Screenshot at 0.33 sec after starting from the same level in each tube is shown.*

The second and third parts start with a screenshot of the sphere in each tube at a certain indication of the length scale. Then, the motion of the sphere in glycerin and in machine oil, respectively, is shown covering a distance of S= 40 cm at time t.

**Use of the Video**: a) Before playing the first part, the student should predict the evolution of the two motions and then check his prediction by following the video. b) The real time motion can be employed to visualize the different terminal velocities acquired by the same sphere falling in different fluids. Moreover, in the second and third part of the video the values of the terminal velocity $u_T = S/t$ of each sphere in the two fluids can be estimated and compared.

2.2 Measurement of sound velocity

In this experiment the velocity of sound in air is estimated by using tuning forks of known frequency that resonate an open – close cylindrical tube of variable length. The wavelength of the sound is determined by making use of the resonance of the air column. The resonance, that is a standing wave in the air column, occurs when the column length L is λ/4, 3λ/4, 5λ/4 where λ is the sound wavelength [ref8, p575].

The first harmonic, where L=λ/4, is indicated by the sudden increase in the intensity of the sound when the column is adjusted to the proper length.

The video (SV) consists of two parts and its duration is 5.20 min. The first part shows the experimental set up and the successive steps to find and measure the length, L, at which the first harmonic occurs for tuning forks with frequencies, ν, 512, 481, 426 and 341 Hz. The sound velocity, υ, is then determined by the equation:

$$υ = ν \cdot λ = 4Lν \qquad (2)$$

In the second part, the procedure to find the first three harmonics for frequencies at 800 and 1000 Hz, using a longer tube is shown (Fig 2).

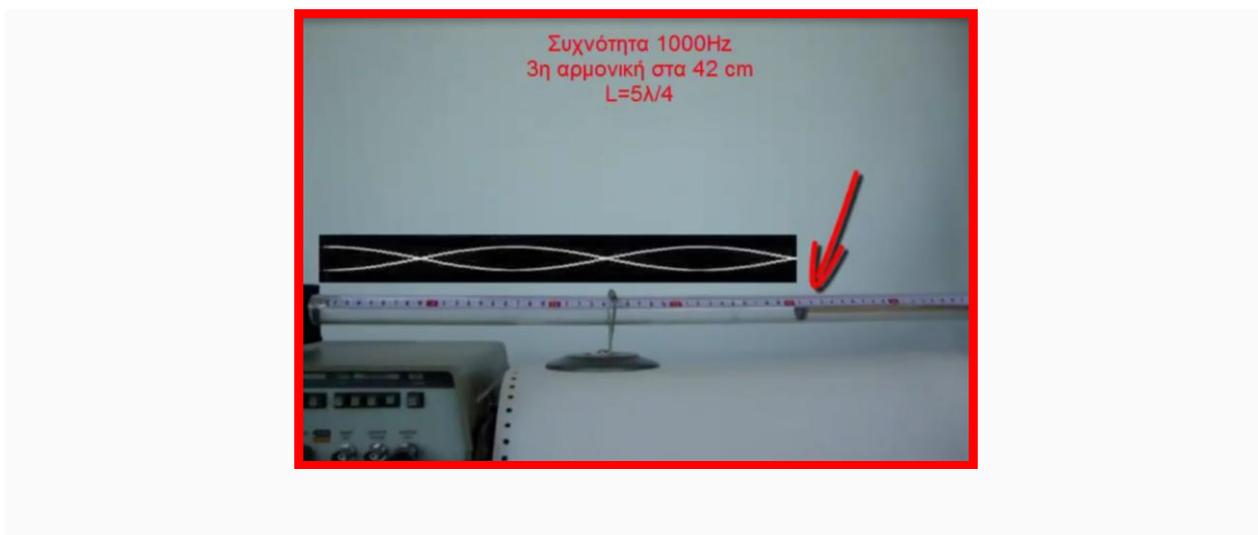

**Figure 2:** *The experimental set up for the measurement of sound velocity and the demonstartion of the three first resonances at a given frequency. The screenshot presents the location of the third resonance (42 cm from the open site of the tube) for the 1000 Hz frequency.*

**Use of the Video**: a) Using the data of the first part, the sound velocity can be estimated graphically, from the slope of the L=L(1/ν) graph, by applying Eq.(2). To get a more accurate value the end correction, which is approximately 0.6 r, can be added (r: is the tube radius). b) The teacher can demonstrate the first harmonic resonance of an air column. In particular, the second part is helpful to explain how the second and third harmonic resonances of the same frequency occur by changing the length L of the tube. Since the corresponding lengths should be λ/4, 3λ/4, and 5λ/4, the students can verify if the column lengths follow the progression 1, 3, 5 and finally calculate the wavelength and velocity of sound. Finally, a comment that in these measurements the end correction is not needed can be given.

2.3 Gamma ray attenuation and measurement of lead attenuation coefficient (GRAC)

In this experiment a $^{60}$Co radioactive source is sited in front of a small detector (Geiger Muller tube) in an appropriate distance and plates of Pb are inserted between them. The counting rate, J, is measured as a function of the thickness of the irradiated material, x, in order to determine the absorption coefficient μ of lead, by using the exponential law:

$$J(x) = J_o\, e^{-\mu x} \qquad (3)$$

where $J_o$, denotes the initial radiation intensity.

The corresponding GRAC video is a 22 min record of the experimental procedure to observe the attenuation of gamma rays passing through lead plates of different thickness. The record processing resulted in a 5.25 min video focusing on a specific range of interest (Fig 3). This range corresponds to the counter indication around 2, 3 and 4 min for four different thicknesses of absorber and for the reference measurement without absorber plates.

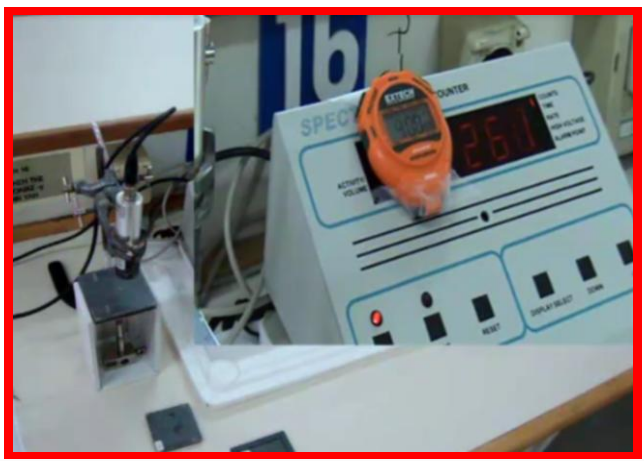

*Figure 3:* The experimental set up for the measurement of the gamma attenuation coefficient. The screenshot shows that 261 counts have been recorded in a time period of 4 min.

Shrinking the experimental procedure to about five minutes results in examining further aspects of the theory and of the experimental procedure.

**Use of Video:** a) During the presentation, the teacher can utilize the full video and focus on the desired range of interest to show the statistical nature of the measurements. Furthermore, students are able to work in groups where each group can elaborate measurements taken in the second, third or fourth min respectively. In doing so, each group analyzes different measurements and the corresponding results can be compared and discussed with respect to the experimental uncertainty. b) By using the counter indication at 4 min for each absorber thickness and after subtracting the background measurement, the attenuation coefficient can be estimated graphically via the relation $\ln J = \ln J_o - \mu x$.

**3. The use of the video GRAC in the physics laboratory**

Having in mind students difficulties and misconceptions (9, 10) in the topic of radioactivity - radiation and related applications a mixed course, of both theory and experimentation, was designed incorporating the video GRAC as well as the use of the real equipment for the background measurement. The lesson was implemented in the physics laboratory of TEI Piraeus during the last semester, March to June 2012.

The objectives of this course were to improve the students' understanding on: the distinction between radioactive and non ionizing radiation, the statistical nature of radioactivity, the exponential law of radioactivity and the meaning of the half life of a nucleus, the interaction between ionizing radiation with matter, the law of exponential attenuation of gamma rays through matter and the meaning of the "half value layer", the natural radioactivity, the beneficial applications of radioactivity as well as the danger from a possible exposure. Finally, students are able to get and analyze the appropriate measurements in order to estimate the gamma attenuation coefficient in lead and the corresponding half value layer.

### 3.1 The objects of the mixed course

The objects include a theoretical presentation and a semi experimental part.

*Theoretical presentation*

- Fundamental issues of radioactivity and the interaction of ionizing radiation with matter focusing on the gamma ray attenuation
- The meaning of half life of a nucleus and the half-value layer of an absorber
- Discussion for the beneficial applications of radioactivity as well as the danger from a possible exposure
- The statistical nature of the counting experiment (using the video- showing the counter indication for min to min)
- Differences between theoretical and experimental data
- The existence of natural radioactivity and its implication on the measurements

*Semi experimental part*

- Active measure of the background radiation (using the same experimental set up as the one used in the video)
- Use of the video to collect measurements dividing students into groups where each group elaborates different data

The course was attended by 51 students from the mechanical engineering department. The 80 % of the students were from integrated and 20% from technical lyceums. Most of the participant students were attending the second semester of their studies whereas all of them successfully attended a physics laboratory during the previous semester. After finishing the teaching – experimenting via video, the students had to analyze their data graphically and present their work while their evaluation was that of a normal laboratory activity. Finally, students were asked to complete a voluntary questionnaire to record their assessment for the use of the video in the lesson and their attitude towards its use as a supplementary educational tool.

## 3.2 Students assessment and attitude towards the use of the video

The evaluation used a five-point response format in which students were asked to rate the degree of their agreement with several statements grouped in three parts. The first part concerned the use of video to collect data (S1, S2), the second one related to the way the laboratory – lesson was presented and executed (S3, S4) and the third part examined their attitude towards the use of video as a supplementary educational tool (S5, S6). The questionnaire was answered by 33 students and the results are summarized in Table1.

| STATEMENTS | STUDENTS RESPONSES % | | |
|---|---|---|---|
|  | Agree | Maybe | Disagree |
| S1. The experience gained by the student from the execution of an experiment, cannot be substituted by taking measurements from a video recorded experiment. | 79 | 12 | 9 |
| S2. Taking measurements from a video recorded experiment could be very useful in case of a very time consuming experiment. | 88 | 12 | 0 |
| S3. The way the laboratory exercise was presented and executed (using a video recorded experiment) gave the opportunity for the discussion and clarification of more issues in comparison to the traditional way of the exercise execution. | 58 | 30 | 12 |
| S4. I believe that the way the laboratory exercise was presented and executed (using a video recorded experiment) was of more benefit to me than if I had executed the experiment together with my colleagues. | 18 | 26 | 54 |
| S5. I believe that the video recorded experiment could be of help to me in the preparation of the experiment. | 79 | 15 | 6 |
| S6. I believe that the use of a video of a real experiment in a lecture, instead of the use of a schematic presentation, could help me to understand better the issues under discussion in the lecture. | 55 | 21 | 24 |

**Table 1:** Students assessment for the use of the video GRAC in the lab and their attitude towards the use of the video as a supplementary educational tool

Since the number of answers is limited, the results strongly agree/disagree and tend to agree/disagree, are grouped together. Regarding students' perceptions for the use of video to collect data, the majority of 79% believe that the experience from the execution of a real experiment cannot be substituted by that from a video recorded experiment. However, 88% of the students agree that the latter one could be useful in case of time consuming experiments. From the students' assessment to the way the laboratory – lesson was presented and executed we conclude that although the majority of the participants (58%) admit that they had the opportunity to discuss and clarify more issues compared to the traditional execution, only 18% believe that the new way of the laboratory exercise was of more benefit to them compared to the usual one. Finally, answers in S5 and S6 show that the majority 79% believes that the video recorded experiment could be helpful for their pre lab preparation and 55% believes that the use of a short video of a real experiment as a supplementary tool could help them to better understand the related issues than using a schematic presentation.

## 4. Summary and conclusions

Three videos of experimental physics[1] were presented that can be integrated easily into a lecture, in the lab or in class activities. The students can use the videos for pre lab preparation to get a first feeling of the real experiment and to be familiar with the tasks and the corresponding calculations involved in the realization of the experiment.

The implementation of the video GRAC in the physics lab was described and the student's assessment and attitude towards the use of the video as a tool in the physics lab and in the classroom were presented and discussed. The answers of the participants showed that students prefer and feel as more beneficial the real experiment instead of the video recorded one. Furthermore, students have a positive attitude towards the use of video as a supplementary educational tool for demonstration and for pre lab preparation.

**References**

[1] Richard E. Berg "Resource Letter PhD-2: Physics Demonstrations", Am. J. Phys. 80 (3), March 2012, http://aapt.org/ajp

[2] http://videolectures.net

[3] http://www.academicearth.org/

[4] Reseed's guide to the 25 best online resources for finding free educational videos http://www.refseek.com/directory/educational_videos.html

[5] http://www.streamingmedia.com/Articles/ReadArticle.aspx?_ArticleID=81019&PageNum=2

[6] Peter B. Kaufman and Jen Mohan "Video Use and Higher Education" http://library.nyu.edu/about/Video_Use_in_Higher_Education.pdf

[7] Douglas Brown, Anne J. Cox " Innovative Uses of Video Analysis",  The Physics Teacher,Vol. 47, pp.145–150 March 2009

[8] Young H., "University Physics", Addison-Wesley (Εκ. Παπαζήση 1990).

[9] Mirofora Pilakouta «Statistical uncertainty in educational experiment on the attenuation of gamma radiation» (Physics session pp 13-20) International Scientific Conference eRA-6 - ISSN-1791-1133, 19- 24 September 2011 TEI Piraeus, Greece

[10]  Mirofora Pilakouta «TEI Piraeus students' knowledge on the beneficial applications of nuclear physics: Nuclear energy, radioactivity – consequences» (Physics session pp 21-27) International Scientific Conference eRA-6 - ISSN-1791-1133 , 19- 24 September 2011 TEI Piraeus, Greece

---

[1] Educational videos : http://eclass.gunet.gr/modules/video/video.php?course=TESTGU220

In case of any problem please contact with mpilak@teipir.gr